# Gell-Mann–Low Function in QCD

## I. M. Suslov

*Kapitza Institute for Physical Problems, Russian Academy of Sciences, ul. Kosygina 2, Moscow, 117973 Russia*
*e-mail: suslov@kapitza.ras.ru*

The Gell-Mann–Low function in QCD $\beta(g)$ ($g = \bar{g}^2/16\pi^2$, where $\bar{g}$ is the coupling constant in the Lagrangian) is shown to behave in the strong-coupling region as $\beta_\infty g^\alpha$, where $\alpha \approx -13$ and $\beta_\infty \sim 10^5$.

Recently, using an algorithm proposed in [1, 2] for summing divergent series of perturbation theory, we determined the Gell-Mann–Low function of $\varphi^4$ theory [1, 2] and QED [3]. Here, this algorithm is applied to QCD, for which previous attempts provided ambiguous results [4].

**1.** Information about all terms of the perturbation series can be obtained by interpolating its first terms with the Lipatov asymptotic behavior [5]. The first four terms in the expansion of the Gell-Mann–Low function for QCD are known in the MS scheme [6]:

$$\beta(g) = \sum_{N=0}^{\infty} \beta_N g^N = \beta_2 g^2 + \beta_3 g^3 + \beta_4 g^4 + \ldots, \quad (1)$$
$$g = \bar{g}^2/16\pi^2,$$

where

$$-\beta_2 = 11 - \frac{2}{3}N_f, \quad -\beta_3 = 102 - \frac{38}{3}N_f,$$

$$-\beta_4 = \frac{2857}{2} - \frac{5033}{18}N_f + \frac{325}{54}N_f^2,$$

$$-\beta_5 = \left[\frac{149753}{6} + 3564\zeta(3)\right] \quad (2)$$

$$- \left[\frac{1078361}{162} + \frac{6508}{27}\zeta(3)\right]N_f$$

$$+ \left[\frac{50065}{162} + \frac{6472}{81}\zeta(3)\right]N_f^2 + \frac{1093}{729}N_f^3.$$

Here, $N_f$ is the number of types of quarks and $\bar{g}$ is the coupling constant in the QCD Lagrangian

$$L = -\frac{1}{4}F_{\mu\nu}^a F_{\mu\nu}^a - \frac{1}{2\xi}(\partial_\mu A_\mu^a)^2$$

$$+ \sum_f \bar{\psi}_f \hat{D} \psi_f + \partial_\mu \bar{\omega}^a(\partial_\mu \omega^a - \bar{g}f^{abc}\omega^b A_\mu^c), \quad (3)$$

$$F_{\mu\nu}^a = \partial_\mu A_\nu^a - \partial_\nu A_\mu^a + \bar{g}f^{abc}A_\mu^b A_\nu^c,$$

$$\hat{D} = i\gamma_\mu(\partial_\mu - i\bar{g}A_\mu^a T^a),$$

where $A_\nu^a$, $\psi_f$, and $\omega^a$ are gluon, quark, and ghost fields, respectively; $T^a$ and $f^{abc}$ are the generators of the fundamental representation and structure constants of the Lee algebra, respectively; $\xi$ is the gauge parameter and subscript $f$ specifies the type of quarks.

**2.** The asymptotic behavior in perturbation theory was discussed for Yang–Mills fields [7–9] and QCD [10, 11], but the results are not sufficiently general. Below, this deficiency will be partially compensated.[1]

The pre-exponential factor of the most general functional integral for QCD involves $M$ gluon, $2L$ ghost, and $2K$ quark fields, i.e.,

$$Z_{MLK} = \int DA D\bar{\omega} D\omega D\bar{\psi} D\psi A(x_1)\ldots A(x_M)$$
$$\times \omega(y_1)\bar{\omega}(\bar{y}_1)\ldots \omega(y_L)\bar{\omega}(\bar{y}_L)\psi(z_1) \quad (4)$$
$$\times \bar{\psi}(\bar{z}_1)\ldots \psi(z_K)\bar{\psi}(\bar{z}_K)\exp(-S\{A,\bar{\omega},\omega,\bar{\psi},\psi\}),$$

where vector indices immaterial for the further consideration are omitted. The substitution $A \longrightarrow B/\bar{g}$ reduces the Euclidean action to the form

$$S\{A,\bar{\omega},\omega,\bar{\psi},\psi\}$$
$$\longrightarrow \frac{S\{B\}}{\bar{g}^2} + \int d^4x\left[\bar{\omega}\hat{Q}\omega + \sum_f \bar{\psi}_f \hat{D}\psi_f\right] \quad (5)$$

---

[1] Our view on the renormalon contributions was formulated in [3]. The existence of renormalon singularities in QCD was neither proven nor disproven, and we shall assume that they are absent.

and the integration over the fermion fields yields

$$Z_{MLK} = (1/\bar{g})^M \int DA B(x_1)\ldots B(x_M)$$
$$\times G(y_1, \bar{y}_1)\ldots G(y_L, \bar{y}_L)\overline{G}(z_1, \bar{z}_1)\ldots \tilde{G}(z_K, \bar{z}_K) \quad (6)$$
$$\times \det\hat{Q}(\det\hat{D})^{N_f} \exp\{-S\{B\}/\bar{g}^2\} + \ldots,$$

where $G$ and $\tilde{G}$ are the Green's functions of the operators $\hat{Q}$ and $\hat{D}$, and ellipsis means terms with other pairings. It is important that $S\{B\}$, $G$, and $\tilde{G}$ are independent of $\bar{g}$. Functional integral (6) is determined by the Yang–Mills action, and the asymptotic behavior of its expansion coefficients in $\bar{g}$ are calculated by the Lipatov method [5]. For the saddle-point configuration, $\bar{g} \sim N^{-1/2}$, where $N$ is the order of perturbation theory. Therefore, each field $A(x_i)$ in the pre-exponential factor in Eq. (4) provides the factor $N^{1/2}$, whereas other fields do not give $N$-dependent factors. The dependence of the expansion coefficients on $N$ is determined according to [10]; it differs from the result for the quark correlation function only by the factor $N^{M/2}$. The $N$th-order contribution to $Z_{MLK}$ has the form

$$[Z_{MLK}]_N \bar{g}^{2N} = \text{const}(16\pi^2)^{-N}$$
$$\times \Gamma\left(N + \frac{M}{2} + 4N_c + \frac{11(N_c - N_f)}{6}\right)\bar{g}^{2N} \quad (7)$$

for even $M$ ($N_c$ is the number of colors), and this expression should be multiplied by the additional factor $\bar{g}N^{1/2}$ for odd $M$.[2]

Using the result for the functional integral and applying the algebra of factorial series [15], one can easily obtain the result for any quantity. Let $F_N \bar{g}^{2N}$ be the $N$th-order contribution to the vacuum integral ($M = L = K = 0$). Then, the general term of asymptotic behavior (apart from a coefficient) has the form $N F_N \bar{g}^{2N}$ for the gluon propagator $\Delta$, $F_N \bar{g}^{2N}$ for the ghost propagator $G$ and quark propagator $G_f$, $N F_N \bar{g}^{2N+1}$ for the gluon–ghost vertex $\gamma_3$ and gluon–quark vertex $\gamma_3^f$, $N^2 F_N \bar{g}^{2N+1}$ for the three-gluon vertex $\Gamma_3$, and $N^3 F_N \bar{g}^{2N+2}$ for the four-gluon vertex $\Gamma_4$. In view of the generalized Ward identities $\Gamma_3 \sim \gamma_3 G$ and $\Gamma_4 \sim \Gamma_3^2 \Delta$, the leading contributions to the asymptotic behaviors of $\Gamma_3$ and $\Gamma_4$ are cancelled, and the invariant charge has the general expansion term $NF_N \bar{g}^{2N+2}$ or $F_N \bar{g}^{2N}$ when determining it from any vertex. The expansion of the $\beta$ function has the same form [5]. Since $g = \bar{g}^2/16\pi^2$, the coefficients of series (1) have the asymptotic behavior

$$\beta_N = \text{const}\,\Gamma\left(N + 4N_c + \frac{11(N_c - N_f)}{6}\right). \quad (8)$$

This result for $N_c = 2$ and $N_f = 0$ agrees with the result obtained in [7].

**3.** Series (1) is nonalternating, and there exists the well-known problem of correct interpretation of the poorly defined Borel integral. In particular, the principal-value prescription for it is not necessarily valid [16]. The definition of the gamma function can be rewritten as

$$\Gamma(z) = \sum_i \gamma_i \int_{C_i} dx\, e^{-x} x^{z-1}, \quad \sum_i \gamma_i = 1, \quad (9)$$

where $C_1, C_2, \ldots$ are arbitrary contours beginning at the origin and tending to infinity in the right half-plane. The Borel transformation of series (1) yields

$$\beta(g) = \sum_i \gamma_i \int_{C_i} dx\, e^{-x} x^{b_0-1} B(gx),$$
$$B(z) = \sum_{N=0}^{\infty} B_N z^N, \quad B_N = \frac{\beta_N}{\Gamma(N+b_0)}, \quad (10)$$

where $b_0$ is an arbitrary parameter. If the Borel transform $B(z)$ has singularities in the right half-plane, contours $C_i$ are no longer equivalent and cannot be reduced to the positive semiaxis, as was possible in Eq. (9). For this reason, the summation result depends on the choice of $\gamma_i$ and $C_i$.[3] We bypass this problem as follows. For the power behavior of the Borel transform at infinity, i.e., when $B(z) \sim z^\alpha$, we have

$$\beta(g) = \beta_\infty g^\alpha, \quad g \longrightarrow \infty$$
$$\text{and } \beta(g) = \bar{\beta}_\infty |g|^\alpha, \quad g \longrightarrow -\infty, \quad (11)$$

where the exact relation between $\beta_\infty$ and $\bar{\beta}_\infty$ depends on the chosen $\gamma_i$ and $C_i$, but $\beta_\infty \sim \bar{\beta}_\infty$ in general case. Therefore, index $\alpha$ can be determined and $\beta_\infty$ can be estimated by summing series (1) for negative values of $g$.

**4.** According to the algorithm developed in [1, 2], the resummation of the alternating series with the coef-

---

[2] The term $M/2$ in the argument of gamma function in Eq. (7) is related to the number of external fields, $4N_c$ is half the number of zero modes, and the term $11(N_c - N_f)/6$ arises because certain zero modes are soft, under more rigorous consideration, and must be nontrivially integrated. For the quark correlation function, Eq. (6) involves divergences, which were removed in [10, 11] by the doubtful method [14]. These divergences are absent for $M \geq 1$.

[3] Results for different $\gamma_i$ and $C_i$ differ by terms proportional to $\exp(-a/g)$, and such nonperturbative contributions must generally be added to the Borel integral. For correctly chosen $\gamma_i$ and $C_i$, these contributions are absorbed by the Borel integral and should not be explicitly taken into account.

ficients behaving asymptotically as $ca^N\Gamma(N+b)$ provides the convergent series with the coefficients

$$U_N = \sum_{K=1}^{N} B_K a^{-K}(-1)^K C_{N-1}^{K-1}, \quad (12)$$

whose behavior for large $N$

$$U_N = U_\infty N^{\alpha-1}, \quad U_\infty = \frac{\bar{\beta}_\infty}{a^\alpha \Gamma(\alpha)\Gamma(b_0+\alpha)} \quad (13)$$

determines the parameters of asymptotic form (11). The coefficient function is interpolated via the formula

$$\beta_N = ca^N N^{\tilde{b}} \Gamma(N+b-\tilde{b})$$
$$\times \left[1 + \frac{A_1}{N-\tilde{N}} + \frac{A_2}{(N-\tilde{N})^2} + \ldots \right] \quad (14)$$

by breaking the series and choosing the coefficients $A_K$ from agreement with Eq. (2). The optimal parameterization of the Lipatov asymptotics with $\tilde{b} = b - 1/2$ is taken [2], and parameter $\tilde{N}$ is used to control the stability of results and to optimize the procedure.

Similar to QED, the parameter $c$ of the Lipatov asymptotics is unknown. In the previous paper [3] it was determined in the course of interpolation. In the case under consideration, such procedure gives large uncertainty in the results, which is not reduced by optimization. So, interpolation was carried out for some predermined $c$ value, which then varied from $10^{-5}$ to $1$.[4] Under this variation, the results change only slightly compared to other uncertainties. Below we present results obtained for $N_c = 3$, $N_f = 0$, and $c = 10^{-5}$.

Fitting $U_N$ by the power law and considering the dependence of $\chi^2$ on $\tilde{N}$, we separate the interval $0.5 \leq \tilde{N} \leq 2.0$, where the $\chi^2$ values are minimal. This procedure determines the set of interpolations consistent with the power behavior of $U_N$. The typical behavior of $\chi^2$ and effective values $U_\infty$ and $\alpha$ as functions of $b_0$ (Fig. 1) indicates that $\alpha \approx -15$. Indeed, $U_\infty$ determined by Eqs. (13) changes sign at $b_0 = -\alpha \approx 15.5$. For this $b_0$ value, $\chi^2$ has a minimum, because the leading contribution $U_\infty N^{\alpha-1}$ vanishes due to the pole of gamma function in Eq. (13), and we have power behavior $U_N \sim N^{\alpha'-1}$ corresponding to the first correction to the asymptotic behavior of $\beta(g)$ [we assume that $\beta(g) = \beta_\infty g^\alpha + \beta'_\infty g^{\alpha'} + \ldots$ for large $g$]. The $\alpha_{\text{eff}}$ value in the first (from large $b_0$) minimum of $\chi^2$ is closest to the exact index $\alpha \approx -15$, because the leading correction to asymptotic form (13) vanishes at $b_0 = -\alpha'$. Different estimations of index $\alpha$ are close to each other when $\tilde{N}$ is close to the optimal value $\tilde{N} = 1.58$ (Fig. 1) and become inconsistent when $\tilde{N}$ go away from its optimal value.

---

[4] Parameter $c$ is equal to the product of the square of the t'Hooft constant $c_H$ in the expression for one-instanton contribution [12, 13] ($c_H^2 \sim 10^{-5}$ and $10^{-4}$ for $N_f = 0$ and 3, respectively) and the dimensionless integral of the instanton configuration. The latter factor can be rather large (characteristic scale is $8\pi^2$).

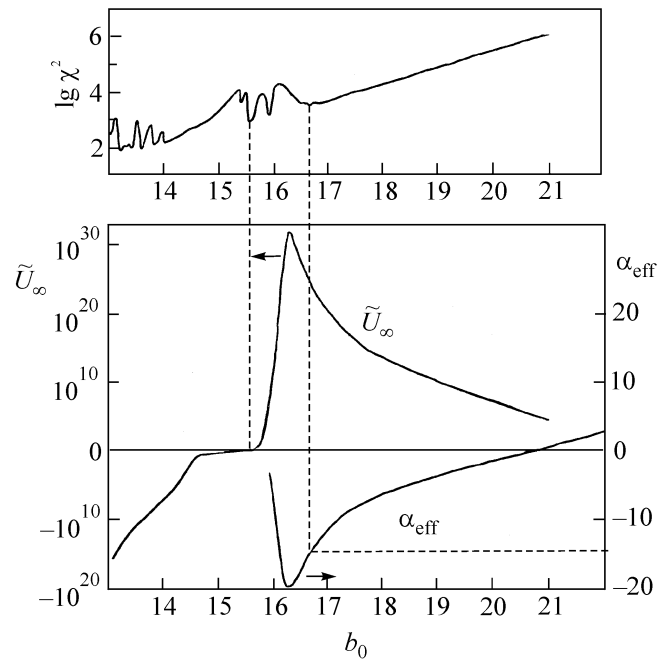

**Fig. 1.** Quantities $\chi^2$, $\alpha_{\text{eff}}$, and $\tilde{U}_\infty = U_\infty \Gamma(b_0+2)$ vs. $b_0$ for the optimal interpolation with $\tilde{N} = 1.58$ and averaging interval $23 \leq N \leq 35$. Function $\tilde{U}_\infty(b_0)$ for $|\tilde{U}_\infty| < 10$ is shown schematically. The minima at $b_0 = 15.4$ and 15.9 are treated as the satellites of the principal minimum at $b_0 = 15.5$, because they, together with the latter minimum, are shifted with varying parameters.

The result for index $\alpha$ cannot immediately be taken as final. First, the large negative index can imitate an exponential. Second, for $\alpha = 0, -1, -2, \ldots$, the leading contribution to the asymptotic behavior of $U_N$ vanishes due to the pole of $\Gamma(\alpha)$ [see Eq. (13)], and the observed result can correspond, e.g., to $\alpha'$ rather than $\alpha$ [2]. In view of these circumstances, we sum a series for the function $W(g) = g^{n_s}\beta(g)$ and increase integer parameter $n_s$ until the observed index $\alpha_W = \alpha + n_s$ becomes positive. The results (Fig. 2a) conclusively demonstrate that we observe a large negative index rather than an exponential. This index is noninteger because in the case $\alpha = -n$ we would observe the behavior shown in the insert. Each point in Fig. 2a is obtained by independent optimization in $\tilde{N}$. The optimal $\tilde{N}$ value decreases mono-tonically with increasing $n_s$. Uncertainty in the results is primarily attributed to the dependence at the lower bound of averaging interval $N_{min} \le N \le N_{max}$. The upper points in Fig. 2a correspond to small $N_{min}$ and to $\chi^2 \sim 10^6$ in the minima. As $N_{min}$ increases, $\alpha$ decreases monotonically until $\chi^2$ reaches a value of $\sim 10^3$ (lower points). With a further increase in $N_{min}$, the pattern of $\chi^2$ minima becomes indistinct, and the uncertainty of the results increases considerably. We admit some further decrease in $\alpha$ until the required values $\chi^2 \sim 10$ are reached, and take this into account in the course of error estimation. Uncertainty in parameter $\bar\beta_\infty$ is of several order of magnitude (Fig. 2b), but the most probable value is $\sim 10^5$, which is consistent with the basic array of data. Thus, we have

$$\alpha = -13 \pm 2, \quad \bar\beta_\infty \sim 10^5 \qquad (15)$$

for $N_f = 0$. One has $\alpha = -12 \pm 3$ and the same most probable value $\bar\beta_\infty$ for $N_f = 3$ (while the total scatter is $\bar\beta_\infty = 1-10^7$). The stability in the results against a change in the summation procedure testifies that their uncertainty is adequately estimated. Some underestimation of the error is possible due to the nonlinear effects [3] and in the case when the asymptotics is reached slowly.

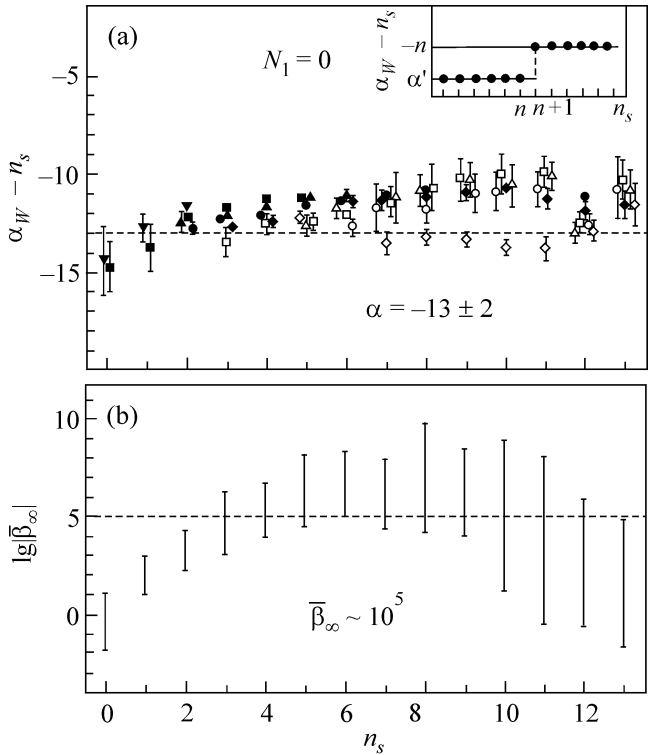

**Fig. 2.** (a) Index $\alpha_W$ obtained by summing the series for function $W(g) = g^{n_s}\beta(g)$ vs. $n_s$ for various averaging intervals $N_{min} \le N \le N_{max}$: (▼) for $N_{min} = 22 + n_s$ and $N_{max} = 35 + n_s$ and (■, ▲, ●, □, △, ○, ◇) for sequentially increasing $N_{min}$ by one unit; (b) parameter $\bar\beta_s$ as a function of $n_s$.

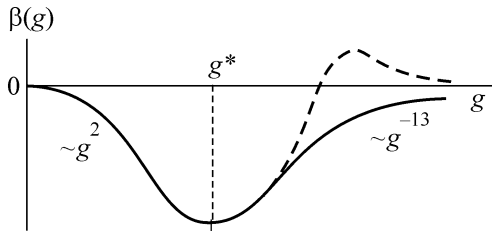

**Fig. 3.**

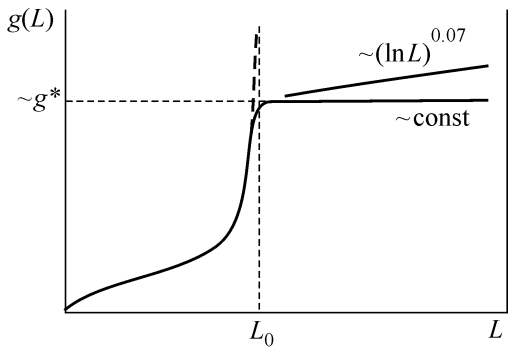

**Fig. 4.**

Large uncertainty in $\beta_\infty$ corresponds to comparatively small uncertainty in the $\beta$ function itself. The characteristic scale where one-loop law $\beta_2 g^2$ is matched with asymptotic behavior (11) appears to be $g^* \sim 2$, and $\bar\beta_\infty$ changes by four orders of magnitude as $g^*$ changes by a factor of two. The sign of $\bar\beta_\infty$ is indeterminate in negative $\alpha_W$ region, because error in $\alpha$ is large and the factor $\Gamma(\alpha)$ in Eq. (13) is alternating, but this sign is definitely negative in positive $\alpha_W$ region (large $n_s$ values). Figure 3 shows (solid line) the behavior of $\beta$ function for $g < 0$ and (dashed line) the analytic continuation to positive $g$ values, where the behavior is qualitatively the same, but the sign of asymptotic function (11) can change.[5] Nevertheless, the behavior of the effective coupling constant as a function of the length scale $L$ is rather definite (Fig. 4). In the one-loop approximation, $g(L)$ has a pole at $L = L_0 = 1/\Lambda_{QCD}$ (dashed line in Fig. 4). For the obtained $\beta$ function (Fig. 3), $g(L)$ increases near $L_0$ up to $\sim g^*$ and then either (for $\beta_\infty > 0$) becomes constant or (for $\beta_\infty < 0$) increases as $(\ln L)^{0.07}$, which is practically indistinguishable from a constant.

In the weak-coupling region, interaction $V(L)$ between quarks is described by he modified Coulomb law $\bar g^2(L)/L$, and the sharp increase in $\bar g(L)$ near $L = L_0$ testifies to the tendency to confinement. In the strong-coupling region, the relation between $V(L)$ and $\bar g(L)$ is

---

[5] In particular, $_\infty = \bar\beta_\infty \cos$ for the principal-value interpretation of the Borel integral.

unknown, but the close in spirit result was obtained by Wilson [17] for the lattice version of QCD:

$$V(L) = \frac{\ln 3\bar{g}^2(a)}{a^2} L, \quad \bar{g}(a) \gg 1 \qquad (16)$$

where $a$ is the lattice constant. From the condition that the result is independent of $a$, the $\beta$ function in the strong-coupling region is estimated as $\beta(g) \sim g \ln g$ [18], which is, however, incorrect. The cross size of the string in the region $a \gg 1/\Lambda_{QCD}$ is equal to $\sim a$, which is considerably higher than its actual physical size $\sim 1/\Lambda_{QCD}$. Therefore, the lattice introduces strong distortions, and there is no reason to expect that the result is independent of $a$. These reasons exist in the region $a \ll 1/\Lambda_{QCD}$, where, however, the coupling constant $\bar{g}(a)$ becomes small, and Eq. (16) does not apply. Thus, Eq. (16) is meaningful only for $a \sim 1/\Lambda_{QCD}$. In the saturation region, $\bar{g}(L) \sim \sqrt{2 \times 16\pi^2} \sim 20$, and, because of a sharp increase in $g(L)$ near $L = L_0$ (Fig. 4), the conditions $a \sim 1/\Lambda_{QCD}$ and $\bar{g}(a) \gg 1$ are compatible, which, likely, justifies the applicability of Eq. (16) to actual QCD.

This work was supported by INTAS (grant no. 99-1070), Russian Foundation for Basic Research (project no. 00-02-17128), and Russian Foundation for Support of Science.

*Translated by R. Tyapaev*